\documentclass[prb,twocolumn,showpacs]{revtex4}
\input epsf
\epsfverbosetrue

\def\ynibc{YNi$_2$B$_2$C}

\usepackage{graphicx}
\usepackage{amsmath}
\usepackage{bm}
\usepackage{epsfig}
\usepackage{subfigure}

\newcommand{\be}{\begin{eqnarray}}
\newcommand{\ee}{\end{eqnarray}}
\newcommand{\nn}{\nonumber}

\newcommand{\mb}[1]{\mathbf{#1}}

\begin{document}
\title{Unconventional superconductivity in
YNi$_2$B$_2$C}
\author{T. R. Abu Alrub and S. H. Curnoe}
\affiliation{Department of Physics and Physical Oceanography,
Memorial University of Newfoundland, St. John's, NL, A1B 3X7, Canada}

\begin{abstract}
We use the semi-classical (Doppler shift) approximation to
calculate magnetic field angle-dependent
density of states and thermal conductivity $\kappa_{zz}$ 
for a superconductor with a quasi-two-dimensional Fermi surface and
line nodes along $k_x=0$ and $k_y=0$.
The results are shown to be in good quantitative agreement 
with experimental results obtained for \ynibc\ (Ref.\
\onlinecite{izawa2002}).
\end{abstract}

\maketitle
\section{Introduction}

YNi$_2$B$_2$C is a  type II superconductor with  a
relatively high transition temperature $T_{\rm c} = 15.6$K.\cite{cava1994}
Although initially thought to be a conventional $s$-wave superconductor,
accumulated evidence soon suggested otherwise.
Power law behaviour in the heat capacity
$C_p/T \propto T^2$ was the first indication that
YNi$_2$B$_2$C is an unconventional superconductor with point nodes in the
gap function.\cite{godart1995,nohara1999}
However, the field dependence of the heat
capacity was found to be $C_p \propto \sqrt H$,
indicative of line nodes.\cite{nohara1999,izawa2001}
The NMR spin relaxation rate $1/T_1$ was measured to be $\propto T^3$ with no
Hebel-Slichter peak\cite{zheng1998}, again consistent with line nodes in the
gap function.
Finally, Raman scattering
showed a peak in the electronic $A_{1g}$ and $B_{2g}$ response,\cite{yang2000}
possibly indicative of a $B_{2g}$ symmetry gap function.
Such a gap function takes the form $\Delta(\bm k) \propto k_xk_y$, 
and thus has symmetry required line nodes along $k_x=0$ and $k_y=0$.\cite{volovik1985,sigrist1991}
In contrast to  these findings, field-angle dependent measurements of
thermal conductivity and specific heat were claimed to be indicative
of point nodes in the gap function.\cite{izawa2002,park2003,matsuda2006}
The reconciliation of these results, and hence
the symmetry of the gap function, remains an important unresolved
issue.

YNi$_2$B$_2$C belongs to the crystallographic space group
I4/mmm (No.\ 139, $D_{4h}^{17}$).\cite{siegrist1994} The lattice is
body-centred tetragonal (bct) with $a=3.526\AA$ and $c=10.543\AA$.\cite{godart1995}
According to symmetry analysis for
$D_{4h}$ crystals,\cite{volovik1985,sigrist1991}
gap functions with line nodes are found only for singlet pairing, while
point nodes are found only for triplet pairing.
Various nodal configurations can occur, depending on the irreducible
representation of $D_{4h}$ by which the superconducting order parameter
transforms.
Nodes in the gap function are normally detected via
quasiparticles (q.p.'s) which appear in the vicinity of gap nodes in $k$-space
as a result of either
finite temperature, impurities, or Doppler shift in the presence of an
applied magnetic field.
In these kinds of measurements, the nodes will be invisible
if there is no
Fermi surface in the direction of the nodes, thus the shape
and connectivity of the
Fermi surface plays an important role.

The Fermi level crosses the 17th, 18th and 19th bands.  The topology of
the Fermi surface is highly sensitive to the precise position of the
Fermi level due to a dispersionless band between the $\Gamma$ and
$X$ points.
Thus different band structure calculations share common features
but the resulting Fermi surfaces have significant differences.\cite{lee1994,singh1996,yamaguchi2004}
Yamaguchi et al.\cite{yamaguchi2004} correlated their results
with de Haas-van Alphen (dHvA) measurements in order to fix the Fermi energy.
The 18th and 19th bands were found to produce closed
Fermi surfaces around various points in the
Brillioun zone, however the 17th band produces a large electron
Fermi surface
multiply connected by necks.
Part of this surface appears as dHvA oscillations perpendicular to the
$c$-axis.  The orbits do not appear to be closed in the $c$ direction; instead they seem to
possess a  two-dimensional character that extends in the $c$ direction,
as evidenced by the upward curvature of
the dHvA frequencies about the $[001]$ direction,
shown in Fig.\ 3 of Ref.\ \onlinecite{yamaguchi2004}.

In the vortex phase of a type II superconductor
q.p.'s may be either
{\em localised} about vortex cores or {\em delocalised}.  It was
shown some time ago that the contribution to the
low-energy density of states in a superconductor with
line nodes comes from {\em delocalised q.p.'s in the vicinity of
the nodes}.\cite{volovik1993}
The delocalised q.p.'s can be
treated with a semi-classical (Doppler shift) approximation;
this approach provided a good description of
field-dependent specific heat and thermal conductivity
of the line node superconductors YBa$_2$Cu$_3$O [\onlinecite{aubin1997}] and CeCoIn$_5$
[\onlinecite{izawa2001}].
However, Volovik's argument does not extend to point node
superconductors.  For point node superconductors,
the semi-classical calculation may still be performed,\cite{tayseer2008}
but, as may be expected,
these results are not in agreement with any experiment
involving putative point node superconductors so far.

In this article, we use the semi-classical
approximation to calculate the field-angle dependent density of states and
thermal conductivity for a superconductor with line nodes and a quasi-2D
Fermi surface, for the purpose of demonstrating that the results
of such measurements on
\ynibc\ are in fact consistent with this scenario, in contrast to
what has been claimed.\cite{izawa2002}

\begin{figure}[ht]
\subfigure[]
{\includegraphics[width=2.7cm]{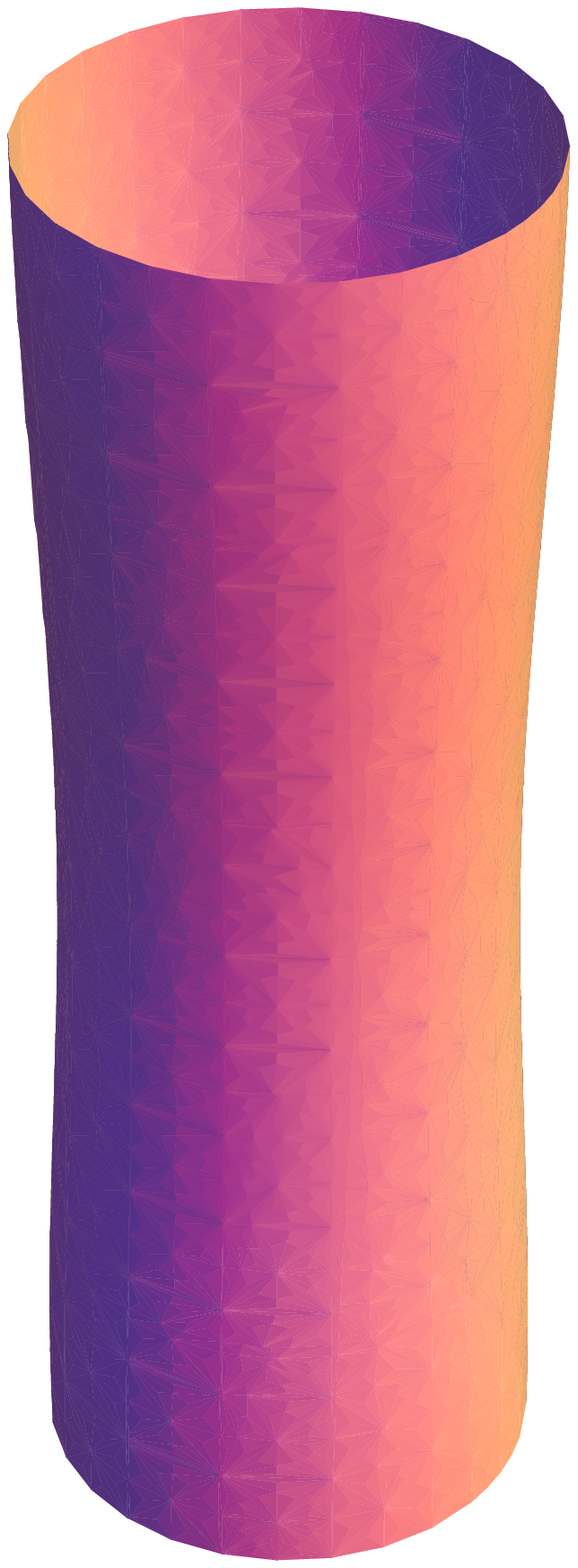}}
\subfigure[]
{\includegraphics[width=2.8cm]{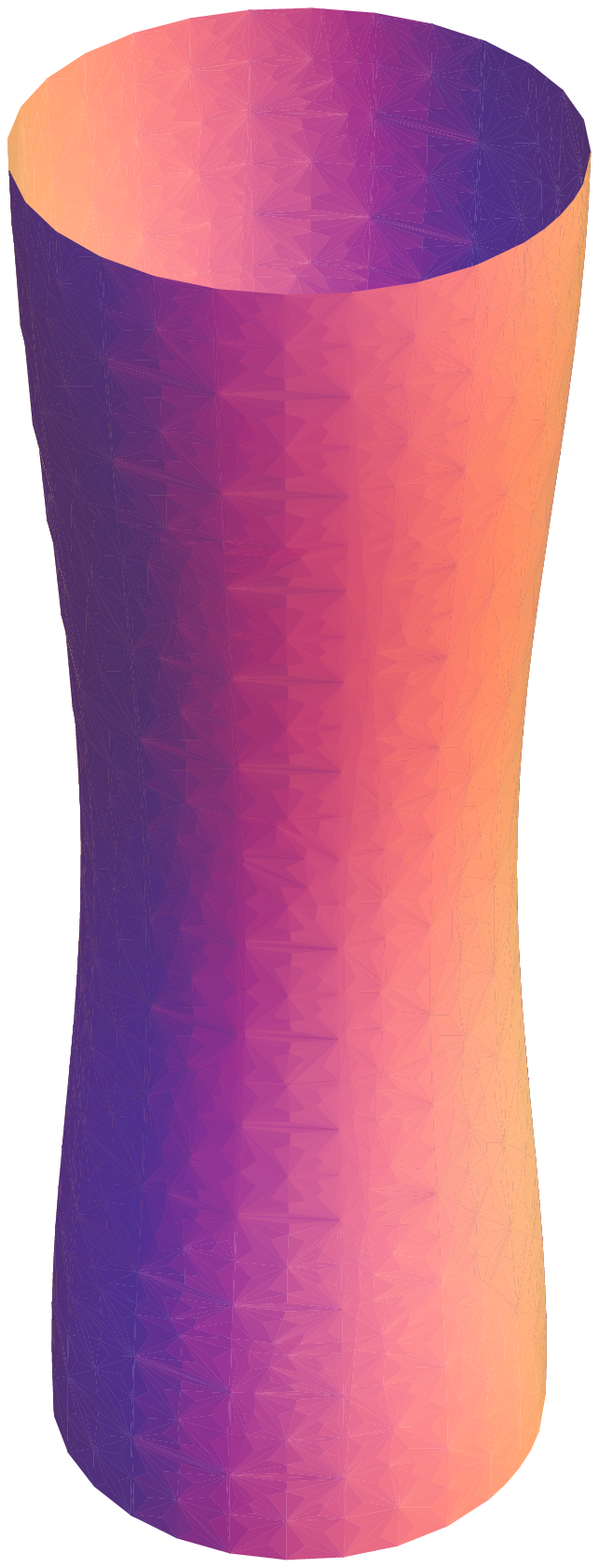}}
\subfigure[]
{\includegraphics[width=2.9cm]{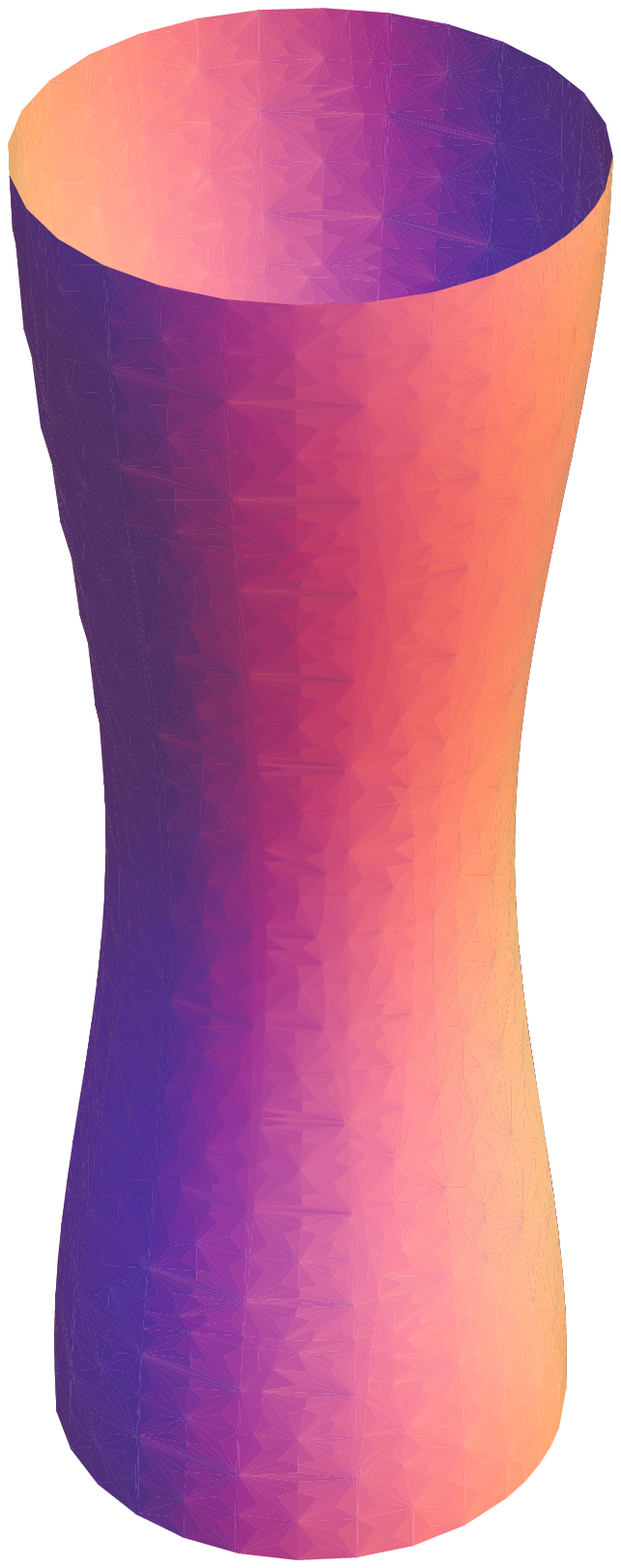}}
\caption{(Colour online) Fermi surfaces for three different values of
$b=\frac{v'_F}{v_F}$.   The surface in c) has the largest value of $b$.\label{figFS}}
\end{figure}

For simplicity, we assume that the Fermi surface has the shape shown in Fig.\
\ref{figFS}, for which the q.p. energy spectrum takes the form
\begin{equation}
\varepsilon({\mb k}) = \frac{k_x^2 + k_y^2}{2m} + \varepsilon_F^{\prime} \cos c k_z
-\varepsilon_F
\end{equation}
where $\varepsilon_F^{\prime}\equiv \frac{k_F^{\prime}c^{-1}}{2m} \ll \varepsilon_F$.
A gap function with $B_{2g}$ symmetry has line nodes
along $k_x=0$ and $k_y=0$.  The Fermi momenta along these nodes (parametrised by $k_z$) are
\begin{eqnarray}
{\bm k}_{F1,3}&=&(0,\pm (2m(\varepsilon_F-\varepsilon_F^{\prime}
\cos{c k_z}))^{1/2}, { k_F'} \sin{c k_z})  \nn \\
& \approx & (0,\pm k_F, k_F^{\prime} \sin c k_z) \label{kf1} \\
{\bm k}_{F2,4}&=&(\pm (2m(\varepsilon_F-\varepsilon_F^{\prime}
\cos{c k_z}))^{1/2},0,k_F' \sin{c k_z})
\nn \\
& \approx & (\pm k_F, 0, k_F^{\prime} \sin c k_z) \label{kf2}. 
\end{eqnarray}
The magnetic field rotates in the $xy$ plane with an
angle $\epsilon$ with respect to the $x$ axis,
\begin{equation}
{\bm H}=H(\cos{\epsilon},\sin{\epsilon},0).
\end{equation}
The supercurrent circulates perpendicular to the
field as a function of the distance $r$ from the vortex core and
winding angle $\beta$,
\begin{equation}
{\bm v_s}(\bm r)=\frac{1}{2 m r}(-\sin{\epsilon} \cos{\beta},
\cos{\epsilon}\cos{\beta},\sin{\beta}) .
\end{equation}
The Doppler shifts associated with each line node are $\alpha_i(\bm r)
= {\bm k}_{Fi}\cdot{\bm v}_s$,
\begin{eqnarray}
\alpha_{1,3}(\bm r)&=&=
\frac{1}{2mr}
[\pm k_F\cos{\epsilon}\cos{\beta}+k_F^{\prime}\sin{\beta}\sin{c k_z}]
\\
\alpha_{2,4}(\bm r)&=&=\frac{1}{2mr}
[\mp k_F\sin{\epsilon}\cos{\beta}+k_F^{\prime}\sin{\beta}\sin{c k_z}]
\end{eqnarray}

\section{Density of states}

\begin{figure}[ht]
\epsfysize=70mm
\epsfbox[-40 350 570 700]{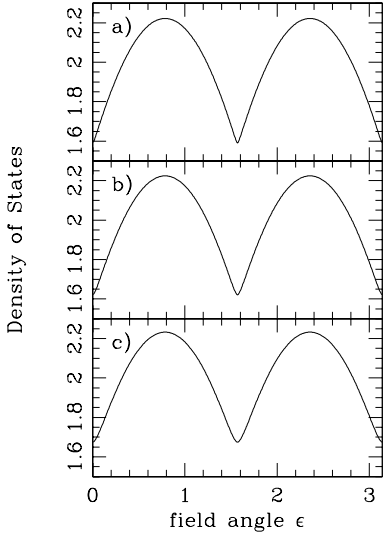}
\caption{Clean limit density of states (the dimensionless expression in the square brackets of
Eq.\ \ref{DOS1}) as a function of rotating field angle $\epsilon$ (in radians) for
$b=0.02$, $0.05$ and $0.10$.
\label{figDOS}}
\end{figure}

In the semi-classical treatment, the argument of the
Green's function $i\omega_n$ is replaced by
$i\omega_n+\alpha$ where $\alpha$ is the Doppler shift.
The quasiparticle energy is
$E(\bm k) = \sqrt{\varepsilon^2(\bm k) + \Delta^2(\bm k)}$
which is $\approx \sqrt{v_F^2k_1^2+v_g^2k_2^2}$ in the vicinity of a node.\cite{durst2000}
Here $k_1$ points in the direction of the node, $k_2$ is perpendicular to $k_1$ in
the $xy$ plane, and the gap velocity is $v_g = \partial \Delta/\partial k_2 |_{\rm node}$.
In the vicinity of the $j$th node, the Green's function takes the form
\begin{equation}
G(\bm k, i{\tilde\omega}_n,{\bm r})=\frac{i{\tilde\omega}_n+\alpha_j(\bm r)+v_F k_1}
{(i{\tilde\omega}_n+\alpha_j(\bm r))^2+v_F^2k_1^2+v_g^2k_2^2}
\end{equation}
where
$i{\tilde\omega}_n=i\omega_n+i\Gamma_0$
and $\Gamma_0$ is the scattering rate at zero energy.
The density of states is
\begin{equation}
N(\omega,\bm r)=-\frac{1}{\pi}\sum_{\bm k}\Im{G(\bm k, \tilde\omega,\bm r)}.
\end{equation}
We divide the volume of integration into four curved cylinder-shaped volumes, each
centred around a line node on the Fermi surface\cite{durst2000} and perform the
integration across the disk spanned by $k_1$ and $k_2$
\begin{eqnarray}
N(0, \bm r) &=&\frac{\Gamma_0}{4\pi^3 v_F v_g}\sum_{j=1}^4\int_{-\pi/c}^{\pi/c}dk_z
\nn\\
&&\times\left[\ln\frac{p_0}{\sqrt{\alpha_j^2(\bm r)+\Gamma_0^2}}+
\frac{\alpha_j(\bm r)}{\Gamma_0}\tan^{-1}\frac{\alpha_j(\bm r)}{\Gamma_0}\right]
\nn\\
\end{eqnarray}
where $p_0$ is the integration cut-off.
In the clean limit $|\alpha_j/\Gamma_0|\gg 1$ the density of states is
\begin{eqnarray}
N(0,\bm r)&=&\frac{1}{4\pi^3 v_F v_g}\int_{-\pi/c}^{\pi/c}dk_z
(|\alpha_1(\bm r)|+|\alpha_2(\bm r)|
\nn\\
&&+|\alpha_3(\bm r)|+|\alpha_4(\bm r)|).
\end{eqnarray}
Averaging over the vortex cross-section, we obtain
\begin{eqnarray}
\left<N(0,\bm r)\right>_H&=&\frac{1}{4\pi^3 v_F v_g}\frac{1}{\pi R^2}
\int_{\xi_0}^Rdr r\int_0^{2\pi}d\beta\int_{-\pi/c}^{\pi/c}dk_z
\nn\\
&&\times(|\alpha_1(\bm r)|+|\alpha_2(\bm r)|+|\alpha_3(\bm r)|
+|\alpha_4(\bm r)|).
\nn\\
\end{eqnarray}
This leads to the result
\begin{eqnarray}
\left<N(0,\bm r)\right>_H&\approx&\frac{8}{\pi^3 v_g c}\frac{1}{\pi R}
\bigg[\sqrt{b^2 + C^2}E\left(\frac{b^2}{b^2+C^2}\right)
\nn\\
&&+\sqrt{b^2+S^2}E\left(\frac{b^2}{b^2+S^2}\right)\bigg]
\end{eqnarray}
where $b=v_F'/v_F$, $c$ is the lattice constant in the $c$ direction,
$C=\cos \epsilon$, $S=\sin\epsilon$  and $E$ is the complete elliptic integral of the second kind.
Using $N_F \sim 1/c v_g \xi_0$ and $\xi_0/R \sim \sqrt{H/H_{c2}}$, we find
\begin{eqnarray}
\left<N(0,\bm r)\right>_H & \sim &  N_F \sqrt{\frac{H}{H_{c2}}}
\bigg[\sqrt{b^2+C^2}E\left(\frac{b^2}{b^2+C^2}\right)
\nn\\
&&+\sqrt{b^2+S^2}E\left(\frac{b^2}{b^2+S^2}\right)\bigg] .
\label{DOS1}
\end{eqnarray}
This function is shown in Fig.\ \ref{figDOS} for various values of $b$.
It is seen that deviations from the perfectly 2D cylindrical Fermi surface leads to a softening
of the cusps in the density of states.

In the dirty limit  $|\alpha_j/\Gamma_0|\ll 1$ we get
\begin{eqnarray}
\left<N(0,\bm r)\right>_H&=&\frac{\Gamma_0}{4\pi^3 v_F v_g}\frac{1}{\pi R^2}
\int_{\xi_0}^Rdr r\int_0^{2\pi}d\beta\int_{-\pi/c}^{\pi/c}dk_z
\nn\\
&\times& \left(4 \ln\frac{p_0}{\Gamma_0}+\frac{\alpha_1^2(\bm r)}{\Gamma_0^2}+\frac{\alpha_2^2(\bm r)}{\Gamma_0^2}
+\frac{\alpha_3^2(\bm r)}{\Gamma_0^2}+\frac{\alpha_4^2(\bm r)}{\Gamma_0^2}\right)
\nn\\
\label{dirtyDOS}
\end{eqnarray}
which produces no oscillations with respect to the rotating field.
\section{Thermal conductivity}
\begin{figure}[ht]
\epsfysize=80mm
\epsfbox[-40 250 570 700]{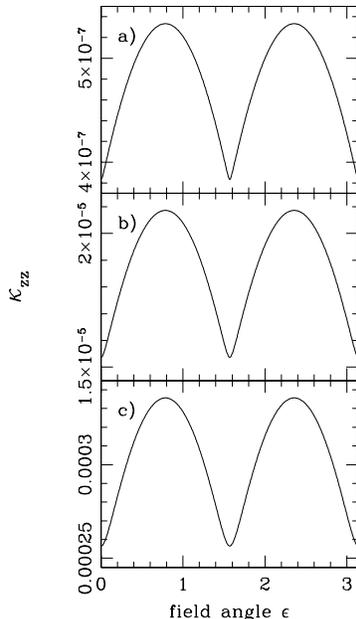}
\caption{Clean limit thermal conductivity $\kappa_{zz}$ (the dimensional expression in the
round brackets of Eq.\ \ref{kappa2}) as a function of
rotating field angle $\epsilon$ (in radians) for $b=0.02$, $0.05$ and $0.10$.
\label{figKappa}}
\end{figure}

The thermal conductivity tensor is given by the Kubo formula, which is expressed in terms of
the imaginary part of the Green's function as
\begin{equation}
\frac{\widetilde{\kappa}(0,{\bm r})}{T}
=\frac{k_B^2}{3}\sum_{\bm k}{\bm v}_F{\bm v}_F\ {\rm Tr}[\Im\widetilde{G}_{ret}(0,{\bm r})\Im\widetilde{G}_{ret}(0,{\bm r})],
\label{kubo}
\end{equation}
where $k_B$ is the Boltzmann constant and
${\bm v_F}$ is the Fermi velocity in the direction of ${\bm k}$.
By again dividing the volume of integration into four regions and
introducing the integration variable $p = \sqrt{v_F^2 k_1^2 + v_g^2 k_2^2}$ we find
\begin{eqnarray}
\frac{\tilde{\kappa}(0,\bm r)}{T}&=&\frac{k_B^2}{3}\frac{1}{(2 \pi^3)v_F v_g}
\int_0^{2\pi}d\phi \int_0^{p_0}dp\,p\int_{-\pi/c}^{\pi/c}dk_z
\nn\\
&&\times \sum_{j=1}^4 ({\bm v}_F{\bm v}_F)_j \frac{2 \Gamma_0^2}
{[(\alpha_j(\bm r)+p)^2+\Gamma_0^2]^2} \\
&=&\frac{k_B^2}{6 \pi^2 v_F v_g}
\int_{-\pi/c}^{\pi/c}dk_z\sum_{j=1}^4 ({\bm v}_F{\bm v}_F)_j
\nn\\
&& \times \left(1+
\frac{\alpha_j(\bm r)}{\Gamma_0}\left(\tan^{-1}\frac{\alpha_j(\bm r)}
{\Gamma_0}-\frac{\pi}{2}\right)\right)
\end{eqnarray}
Using (\ref{kf1}, \ref{kf2}),
in zero magnetic field we get
\be
\frac{\tilde{\kappa}(0,0)}{T}&=&
\frac{2 k_B^2v_F}{3 \pi c v_g}\left(\begin{tabular}{lclclc}
1&&0&0\\
0&&1&0\\
0&&0&$(\frac{v_F'}{v_F})^2$
\end{tabular}\right).
\ee
In a finite magnetic field, terms linear in the Doppler shift will vanish upon
integration.
So the magnetic part of the thermal conductivity is
\begin{eqnarray}
\frac{\delta\tilde{\kappa}(0,\bm r)}{T}&=&\frac{k_B^2}{6 \pi^2 v_F v_g}
\int_{-\pi/c}^{\pi/c}dk_z\sum_{j=1}^4 ({\bm v}_F{\bm v}_F)_j
\nn\\
&&
\frac{\alpha_j(\bm r)}{\Gamma_0}\tan^{-1}\frac{\alpha_j(\bm r)}
{\Gamma_0}
\label{thermal1}
\end{eqnarray}
In the clean limit, this reduces to
\begin{eqnarray}
\frac{\delta\tilde{\kappa}(0,\bm r)}{T}&=&\frac{k_B^2}{12 \pi v_F v_g}
\int_{-\pi/c}^{\pi/c}dk_z\sum_{j=1}^4 ({\bm v}_F{\bm v}_F)_j
\frac{|\alpha_j(\bm r)|}{\Gamma_0}
\nn . \\
\end{eqnarray}
The integrand is
\begin{widetext}
\begin{equation}
\frac{v_F^2}{\Gamma_0}
\left(\begin{array}{ccc}
|\alpha_2|+|\alpha_4|&0&\frac{v_F'}{v_F} (|\alpha_2|-|\alpha_4|)\sin{c k_z}\\
0&|\alpha_1|+|\alpha_3|&\frac{v_F'}{v_F}(|\alpha_1|-|\alpha_3|)\\
\frac{v_F'}{v_F}(|\alpha_2|-|\alpha_4|)\sin{c k_z}&\frac{v_F'}{v_F}(|\alpha_1|-|\alpha_3|)\sin{c k_z}&(\frac{v_F'}{v_F})^2(|\alpha_1|+|\alpha_2|+|\alpha_3|+|\alpha_4|)\sin^2{c k_z}
\end{array}\right).
\end{equation}
\end{widetext}
The off-diagonal components vanish in the vortex average, and the diagonal
components are
\begin{eqnarray}
\frac{\langle\delta\kappa_{xx}\rangle_H}{T} & = & \frac{4}{3\pi^2}\frac{k_B^2}{R \Gamma_0} \frac{v_F^2}{v_gc}
\sqrt{b^2+S^2}E\left(\frac{b^2}{b^2+S^2}\right) \\
\frac{\langle\delta\kappa_{yy}\rangle_H}{T} & = & \frac{4}{3\pi^2}\frac{k_B^2}{R \Gamma_0} \frac{v_F^2}{v_gc}
\sqrt{b^2+C^2}E\left(\frac{b^2}{b^2+C^2}\right)  \\
\frac{\langle\delta\kappa_{zz}\rangle_H}{T} &= &  \frac{4}{9\pi^2}\frac{k_B^2}{R \Gamma_0} \frac{v_F^{2}}{v_gc}
\left(\sqrt{b^2+S^2}\left[-S^2K\left(\frac{b^2}{S^2+b^2}\right) \right.
\right. \nonumber \\
& & \hspace{.7in} \left. +(2b^2+S^2)E\left(\frac{b^2}{S^2+b^2}\right)\right] \nonumber \\
& & \hspace{.4in} + \sqrt{b^2+C^2}\left[-C^2K\left(\frac{b^2}{C^2+b^2}\right)\right.
\nonumber \\
& & \hspace{.7in} \left. \left. +(2b^2+C^2)E\left(\frac{b^2}{C^2+b^2}\right)\right]\right)
\label{kappa2}
\end{eqnarray}
where $K$ is the complete elliptic integral of the first kind.
$\kappa_{zz}$ is plotted in Fig.\ \ref{figKappa} for different values of $b$.  In the
limit $b\rightarrow 0$ the cusps are sharp, however the oscillation
amplitude goes to zero.  The oscillation amplitude increases exponentially 
with $b$.

In the dirty limit, (\ref{thermal1}) reduces to
\begin{eqnarray}
\frac{\delta\tilde{\kappa}(0,\bm r)}{T}&=&\frac{k_B^2}{6 \pi^2 v_F v_g}
\int_{-\pi/c}^{\pi/c}dk_z\sum_{j=1}^4 ({\bm v}_F{\bm v}_F)_j
\frac{\alpha_j^2(\bm r)}{\Gamma_0^2}
\nn\\
\end{eqnarray}
Again, the off-diagonal components vanish in the vortex average, and the
diagonal elements are
\begin{eqnarray}
\left\langle \frac{\delta\kappa_{xx}(0,\bm r)}{T} \right\rangle_H & = & \frac{k_B^2v_F^3}{12 \pi v_g}\log(R/\xi_0)
(2 C^2 + b^2) \\
\left\langle \frac{\delta\kappa_{yy}(0,\bm r)}{T} \right\rangle_H & = & \frac{k_B^2v_F^3}{12 \pi v_g}\log(R/\xi_0)
(2 S^2 + b^2) \\
\left\langle \frac{\delta\kappa_{zz}(0,\bm r)}{T} \right\rangle_H & = & \frac{k_B^2v_F^3}{12 \pi v_g}\log(R/\xi_0)
(1+3b^2/2)
\end{eqnarray}
Similar to the dirty limit density of states (\ref{dirtyDOS}), there are no rotating field-dependent
oscillations in the dirty limit of $\kappa_{zz}$.


\section{Discussion and Conclusions}

The topology of the true Fermi surface of \ynibc\ shown in Ref.\
\onlinecite{yamaguchi2004} is difficult to discern, however the validity of our
calculation only requires that the Fermi surface exists at the positions of the nodes and
spans all or most of the Brillioun zone in the $c$ direction with a slight
curvature characterised by the parameter $b$.
The main point is that the cusp features observed in the field angle-dependent
heat capacity\cite{park2003} are a feature of {\em line nodes} and the
cusp features observed in the field angle-dependent thermal conductivity 
are a
feature of {\em line nodes on a quasi-2D Fermi surface}.
In contrast, the semi-classical (Doppler shift)
treatment of point nodes (which may not even be valid)
is insensitive to Fermi surface topology and produces neither cusp
features {\em nor} four-fold oscillations.\cite{tayseer2008}

Using $R \approx 4\times 10^{-8}$ m (for a 1 T field), 
$T=0.56$ K, 
$E_F = 9$ Ry,\cite{yamaguchi2004}
$v_F \approx 3\times 10^{7}$ m/s,
gap maximum $\Delta_0 = v_g \hbar k_F = 30$ K and
scattering rate 
$\Gamma_0 = 1$ K
leads to an estimate of the prefactor 
in (\ref{kappa2}) of
$\frac{4 T}{9\pi^2} \frac{k_B^2}{R\Gamma_0}\frac{v_F^2}{v_g c}
\approx 10^4$ W/Km.
The experimentally observed oscillation amplitude is
$\approx 2\times 10^{-3}$.
Comparing with the oscillation amplitudes shown in 
Fig.\ \ref{figKappa}, one may deduce that the value
of $b$ is approximately 0.02.  Such a small value of $b$ produces sharp
cusps in the field angle-dependent $\kappa_{zz}$ oscillations and is 
therefore fully consistent with  experiment.

Thus the most straight-forward model that best describes accumulated
observations on \ynibc\ is that the superconducting order parameter is
$\Delta \sim k_xk_y$, which belongs to the irreducible representation
$B_{2g}$ of the point group $D_{4h}$, with associated line nodes
along $k_x=0$ and $k_y=0$.

\end{document}